\documentclass[pdflatex,sn-mathphys-num]{sn-jnl}

\usepackage{graphicx}%
\usepackage{multirow}%
\usepackage{amsmath,amssymb,amsfonts}%
\usepackage[normalem]{ulem}
\usepackage{amsthm}%
\usepackage{mathrsfs}%
\usepackage[title]{appendix}%
\usepackage{xcolor}%
\usepackage{textcomp}%
\usepackage{manyfoot}%
\usepackage{booktabs}%
\usepackage{algorithm}%
\usepackage{algorithmicx}%
\usepackage{algpseudocode}%
\usepackage{listings}%

\begin{document}

\title[Fundamental Klein-Gordon equation from stochastic mechanics in curved spacetime.]{Fundamental Klein-Gordon equation from stochastic mechanics in curved spacetime.}


\author*[1]{\fnm{Eric S.} \sur{Escobar-Aguilar}}\email{e.escobar@xanum.uam.mx}

\author[2]{\fnm{Tonatiuh} \sur{Matos}}\email{tonatiuh.matos@cinvestav.mx}
\equalcont{These authors contributed equally to this work.}

\author[1]{\fnm{J.I.} \sur{Jimenez-Aquino}}\email{ines@xanum.uam.mx}
\equalcont{These authors contributed equally to this work.}

\affil*[1]{\orgdiv{Departamento de F\'isica}, \orgname{Universidad Aut\'onoma Metropolitana Iztapalapa}, \orgaddress{\street{San Rafael Atlixco 186}, \city{CDMX}, \postcode{09340}, \state{Ciudad de M\'exico}, \country{M\'exico}}}

\affil[2]{\orgdiv{Departamento de Física}, \orgname{Centro de Investigación y de Estudios Avanzados del Instituto Politécnico Nacional}, \orgaddress{\street{Av. Instituto Politécnico Nacional 2508}, \city{CDMX}, \postcode{07360}, \state{Ciudad de M\'exico}, \country{M\'exico}}}


\abstract{This work presents an alternative approach to obtain the quantum field equations in curved spacetime, considering that sufficiently small particles follow stochastic trajectories around geodesic. Our proposal is based on a stochastic differential equation in which the noise term experienced by the quantum particles is a consequence of the stochastic background in spacetime. This fact allows the particles to describe erratic movements and locally the universe exhibits characteristics akin to a lake with gentle ripples rather than a flat unyielding surface. Building upon this foundational understanding, we investigate the influence of this background on quantum-scale particles without considering the metric to be stochastic, rather we let test particles move randomly around the geodesic of macroscopic particles. Their behavior aligns with solutions to the Klein-Gordon (KG) equation specific to this curved spacetime. As the KG equation, in its non-relativistic limit within a flat spacetime, reduces to the Schrödinger equation, consequently, we propose a compelling connection: the Schrödinger equation may emerge directly from a spacetime lacking local smoothness.}




\maketitle

\section{Introduction}
Quantum Mechanics (QM) and General Relativity (GR) stand as towering achievements in 20th-century physics, fundamentally reshaping our understanding of the cosmos. The inception of the uncertainty principle within QM revolutionized our perception of the microworld, yet paradoxes within QM persist, and reconciling GR with QM remains unsolved. This paradox is accentuated by the absence of an accepted quantum theory of gravitation, recently an alternative approach was presented and considered quantum field theory coupled to classical spacetime \cite{ PhysRevX13041040}. Consequently, the theory must exhibit an essential stochastic nature, and in Oppenheim's theory, this coupling is studied via a master equation for the density matrix $\rho$ without the necessity to consider a stochastic metric.

In the present work, we start from the hypothesis that spacetime is classical but exhibits stochastic behavior. We show that this hypothesis implies that particles follow trajectories that satisfy the Klein-Gordon (KG) equation, or in its Newtonian limit, the Schr\"odinger equation. This gives a new interpretation of QM where the Schr\"odinger equation is a consequence of the stochastic structure of spacetime.
The sources of the stochasticity in the theory can be attributed to the existence of Gravitational Wave Background (GWB), originating from events like the Big Bang, inflation, and cosmic transitions, permeates the fabric of spacetime, akin to gentle ripples on a cosmic lake. Recent observations underscore this existence, although identified signals primarily reside within nanohertz frequencies \cite{NANOGrav:2023gor, Reardon:2023gzh, Xu:2023wog, antoniadis2023second}. Nonetheless, compelling theories tie the GWB to the accelerated expansion of the universe \cite{Matos:2021jef} and align with cosmological observations of the universe, the Cosmic Microwave Background Radiation, and the profiles of the Mass Power Spectrum \cite{Matos:2022jzf, Matos:2023qwx}.

GR delineates spacetime as a smooth differentiable manifold, locally flat and enabling geodesic trajectories between points. However, our work delves into a paradigm where the spacetime is no longer locally flat and obstructs such deterministic paths for sufficiently small particles. Instead, these particles are guided by stochastic trajectories due to spacetime fluctuations, this could be achieved by considering two approaches: one assuming a stochastic metric with very small fluctuations and the other one, considering small test particles traveling around the geodesics with a stochastic term added to the trajectory. The former is usually called stochastic gravity \cite{ Hu:1999mm, StochasticGravity,Okon:2013lsa}, and the latter could be studied considering a stochastic differential equation, or using an equation associated with the probability density. 

While stochastic gravity framework generalizes semiclassical gravity by incorporating quantum fluctuations of the stress-energy tensor into the Einstein equations via a stochastic noise term that describes the fluctuations of quantum matter fields in curved spacetime, leading to the Einstein-Langevin equation, in the other hand our research considers a stochastic differential equation for the trajectories and demonstrates that such trajectories, within a curved spacetime, lead to the KG equation. Here, the norm and phase of the complex KG function respectively conduct to stochastic and hydrodynamic velocities. Remarkably, in its flat non-relativistic limit, the KG equation converges to the Schrödinger equation, suggesting a profound link between the spacetime fluctuations and the emergence of quantum principles.
Both theories, although dealing with stochastic terms, are different. The first one incorporates stochasticity into the spacetime metric via the noise kernel in the Einstein-Langevin equation. The presented theory introduces stochasticity in the particle trajectories, which affects particle motion but leaves the spacetime metric deterministic.

Exploring this notion further, historical works like Marshall's on Random Electrodynamics \cite{1963} and Boyer's treating 
the Zero-Point Radiation Field \cite{boyer1968quantum}  postulates that the source of stochasticity arises from the interaction of particles with the electromagnetic zero-point radiation field, which fills space with electromagnetic energy.  This work has been used to expose non-relativistic quantum electrodynamics in the Weyl–Wigner representation \cite{santos2022analogy}. On the other hand, Nelson's stochastic treatment of Newton's second law elucidates connections between quantum mechanics and stochastic processes \cite{Nelson_1966}. These approaches are consistent with the standard rules of quantum mechanics, such as the Born rule and the uncertainty principle. Some advantages of the stochastic interpretation are also discussed, for instance, a natural explanation for the collapse of the wave function of the Schrödinger equation, and some of the conceptual problems of quantum mechanics \cite{pavon1999derivation}. Both approaches are suitable for the investigation of the present work, and we mainly follow the work of  L. de la Pe\~na and A. M. Cetto in \cite{Emergingquantum}. Moreover, recent endeavors including stochastic gravity \cite{haba2022quantum,namsrai1991stochastic} and Nottale's scale relativity, showcase promising lines of investigation towards understanding the quantum behavior in curved spacetime \cite{chavanis2017derivation} and its relevance to dark matter halos \cite{chavanis2018derivation}.

In our present and main investigation, we explore relativistic stochastic mechanics within a fluctuating curved spacetime. Our formulation employs Markovian stochastic differential equations, allowing quantum particles to follow stochastic trajectories around geodesic. Consequently, we derive the KG equation in curved spacetime leading to a generalized Schrödinger equation, unveiling the close relationship between quantum fluctuations in spacetime and fundamental quantum equations \cite{chavanis2017covariant}. In this direction, the hydrodynamic version of the KG equation will be the middle point for both theories.
Moreover, by adopting the hydrodynamic representation of quantum equations \cite{Matos_2019}, we uncover insights into quantum fluids and particle behavior in external potentials \cite{chavanis2017covariant}. This representation is explored in section II.

In section III the derivation of continuity, Hamilton-Jacobi, and Euler equations from a stochastic standpoint, elucidating the multifaceted interplay between quantum mechanics, stochasticity, and curved spacetime. And consequently the success of Nelson's stochastic mechanics to achieve our goal. Section IV is dedicated to concluding and discussing the main results of this approach.

\section{Field equations}

The groundwork for the proposed formalism is the hydrodynamic form of the Klein-Gordon equation in curved spacetime with an arbitrary scalar field potential given by $V=2m^2\mathcal{A}\Phi\Phi^\dagger$ that is minimally coupled to a gauge vector field through electromagnetic interaction presented in \cite{Matos_2019}\footnote{As in the cited work, in this section we use natural units, that is, $c=\hbar=\epsilon_{0}=\mu_{0}=1.$}.
\begin{eqnarray}
\square_{E}\Phi-\frac{d V}{d\Phi^{\dagger}}&=&0,\label{eq:KG}\\
\nabla_\nu F^{\nu\mu}&=&J^\mu\label{eq:Maxwell}\\
J_\mu&=&\frac{i}{2m^2}[\Phi(\nabla_\mu-ieA_\mu)\Phi^\dagger-\Phi^\dagger(\nabla_\mu+ieA_\mu)\Phi]\label{eq:flux}
\end{eqnarray}
Here $m$ is a mass parameter, $\mathcal{A}$ is convenient parametrization of the scalar field potential $V$, and $J_\mu$ is the scalar field flux. The scalar field $\Phi=\Phi(x^\mu)$ is a complex function of the coordinates $x^\mu$ and $\Phi^\dagger$ its complex conjugate.
The KG equation is expanded using coordinates in a four-dimensional manifold whose geometry is determined by a metric $g$, and that acts as the curved physical spacetime.
The d’Alembert operator of interest is of the form
\begin{equation}
    \Box_{E}=\left( \nabla^{\mu}+i e A^{\mu}\right)\left( \nabla_{\mu}+i e A_{\mu}\right),
\end{equation}
with $\nabla^\mu$ the covariant derivative associated to the metric $g$, $e$ is the charge of the particle and $A_{\mu}$ the gauge vector field associated with the Maxwell $4-$potential with Maxwell tensor $F_{\mu\nu}=\nabla_\mu A_\nu-\nabla_\nu A_\mu$.

We will work in an $ADM$ 3+1 foliation of the spacetime (from  R. Arnowitt, S. Deser, and C. W. Misner) that considers 3 space-like hypersurfaces and an evolution parameter $t$ that describes the temporal evolution of these 3 space-dimensions. This formalism allows studying spacetime in terms of its spatial geometry and temporal dynamics. The metric is given by
\begin{equation}
ds^{2}=-N^{2}d(ct)^{2}+\gamma_{ij}\left(dx^{i}+N^{i}dt\right)\left(dx^{j}+N^{j}dt\right), \label{ADM}
\end{equation}
where $N^i$ describes the relative displacement of spatial coordinates between successive hypersurfaces, $N$ is the lapse function, which describes how time flows between neighboring hypersurfaces, and $\gamma_{ij}$ defines the intrinsic geometry of the 3-dimensional hypersurface. This allows us to express the field equations as a set of evolution equations for the metric on each hypersurface and a set of constraint equations that must be satisfied at every point on each hypersurface \cite{Alcubierre_2008}.

The KG equation exhibits the remarkable feature of permitting localized, non-dispersive solutions under specific conditions, making it particularly relevant for physical phenomena such as scattering. However, Derrick’s theorem establishes that in flat spacetime, static and localized scalar field solutions are inherently unstable \cite{derrick1964comments}. This issue can be addressed by applying a harmonic decomposition to the complex scalar field, incorporating the time dependence as a phase factor. The field $\Phi$ is expressed as:

\begin{equation}
    \Phi(x^{0}, \mathbf{x}) = \Psi(x^{0}, \mathbf{x}) e^{-i\omega_{0}x^{0}}, \label{eq:KG harmonic}
\end{equation}
where $\omega_{0}$ represents either the mass or frequency of the particle for massless particles, $\Psi$ denotes the amplitude, and the particle density is given by $n(t, \mathbf{x}) = |\Phi|^2 = |\Psi|^2$. This method removes the staticity of the field while preserving the static nature of spacetime, allowing soliton-like solutions to exist \cite{rosen1966existence}.

The hydrodynamic representation to express the field equations derived from KG recall the Madelung transformation \cite{Madelung1927QuantentheorieIH}. This transformation shows the equivalence between the Schrödinger equation and the Euler equation for an irrotational fluid, adding a quantum potential. The boson gas $\Phi$ can be viewed as a real fluid, which is described by quantum Euler equations. Then the solution (\ref{eq:KG harmonic}) can be decomposed into its hydrodynamic form by taking the amplitude $\Psi=\sqrt{n} e^{i\theta}$ as the Madelung transformation, considering $x^{0}=t$ the evolution parameter of the metric (\ref{ADM}), this results in  $\Phi(t, \mathbf{x})$ taking the form
\begin{equation}
\Phi(t, \mathbf{x}) = \sqrt{n} e^{i(S - \omega_0 t)}. \label{eq:madelung}
\end{equation}
Here, the complex scalar field is decomposed into a density $n(t, \mathbf{x})= |\Psi(t, \mathbf{x})|^2$ is the probability density (interpreted as the fluid density in hydrodynamic terms), and a phase function $\theta(t, \mathbf{x}) = S(t, \mathbf{x}) - \omega_0 t$, where $S(t, \mathbf{x})$ encodes the geometric information of the system.

In the Madelung formalism, $S(t, \mathbf{x})$ plays the role of the action, and the definition of velocity aligns with Hamilton-Jacobi's theory in classical mechanics, where the velocity is proportional to the gradient of the action. Then the momentum per unit mass $m,$ or velocity corresponds to
\begin{equation}
    v^\mu \equiv \frac{1}{m}\left( \nabla^\mu S + eA^\mu \right),
\end{equation}since in the Madelung transformation (\ref{eq:madelung}) the phase function includes $\omega_0 t$ the generalized velocity is defined as
\begin{equation}
\pi^{\mu}= \dfrac{1}{{m}} \left(\nabla^{\mu}
\theta +e A^{\mu}\right)=v^{\mu}- \dfrac{\omega _{0}}{{m}}\nabla^{\mu}t.
\label{eq:generalized velocity}
\end{equation}

Further following the Madelung formalism, by substituting (\ref{eq:madelung}) in (\ref{eq:KG}) the KG equation splits into two separate equations, one corresponding to the real part
\begin{equation}
\nabla_\mu \nabla^\mu \sqrt{n} -\sqrt{n}\nabla_\mu \theta\nabla^\mu \theta - 2 e \sqrt{n} A^\mu \nabla_\mu \theta - e^2 \sqrt{n} A^\mu A_\mu - 2m^2\sqrt{n}\mathcal{A}=0,
\end{equation}where the potential is written as $V=2m^{2}\mathcal{A}\Phi\Phi^{\dagger},$ and for the imaginary part
\begin{equation}
    \nabla_\mu\sqrt{n}\left( 2\nabla^\mu \theta + eA^\mu \right)   + \sqrt{n} \nabla_\mu \nabla^\mu \theta + e \nabla_\mu (\sqrt{n} A^\mu ) = 0.
\end{equation}Now using the generalized velocity (\ref{eq:generalized velocity}), after some algebra, the imaginary part results in the continuity equation
\begin{equation}
\nabla_{\mu}\left(nv^\mu\right)-\frac{\omega_{0}}{m} \left(\nabla^{0}n+n\square t\right)=0,\label{Continuity Tona}
\end{equation}
while for the real part the Hamilton-Jacobi is recovered
\begin{equation}
v_{\mu}v^{\mu}-2\frac{\omega_{0}}{m} v^{0} - \frac{\omega_{0}^{2}}{m^{2}N^{2}} + 2\mathcal{A} -\frac{\square\sqrt{n}}{m^{2}\sqrt{n}}=0,\label{H-J Tona}
\end{equation}
where $V^Q=\frac{\square\sqrt{n}}{m^{2}\sqrt{n}},$ denotes the quantum potential (See reference \cite{Matos_2019} for a complete derivation.). These two equations represent the hydrodynamic version of the KG equation written in the Madelung (or hydrodynamic) variables $n$ and $v^{\mu}$.

To transition toward a stochastic formalism, Eqs. (\ref{Continuity Tona}) and (\ref{H-J Tona}) shall be reformulated in terms of $\pi^\alpha$ and $n$:
\begin{equation}\label{eq:Continuidad}
\nabla_\alpha\left(n \pi^\alpha\right)=0
\end{equation}
and
\begin{equation}\label{eq:Navier}
    \frac{1}{2}\pi_\alpha\pi^\alpha+\mathcal{A}-\frac{1}{2m^2}\frac{\Box\sqrt{n}}{\sqrt{n}}=0.
\end{equation}
Observe that Eqs. (\ref{eq:Continuidad}) and (\ref{eq:Navier}) are the KG equation written in the variables $\pi^\alpha$ and $n$.


\section{Stochastic Quantum Mechanics}
The assumptions considered to establish the stochastic mechanism in curved spacetime are the following:

1) The quantum particles interact with the universally present stochastic spacetime background, analogous to Brownian motion in classical physics. This interaction is not merely a small perturbation but is fundamental in describing the quantum behavior.
By associating stochasticity with the spacetime background itself it is universally applicable to all particles and fields, differing from the stochastic electrodynamics. Also working with a deterministic metric, the submitted theory retains compatibility with standard relativistic principles while incorporating stochasticity at the particle level, avoiding mathematical complications like solving Einstein-Langevin equations.

2) The interaction between a quantum particle and the background radiation is not fully knowable in detail. Instead, only its statistical properties are considered, like the statistical properties governing Brownian motion.

To clarify our proposal, we start with the definition of the 4-velocity $\mathcal{U}^\mu$, and we add a stochastic term to the definition
\begin{equation}
   \frac{d x^\mu}{d\tau}=\mathcal{U}^\mu + \sqrt{2\sigma}\, \xi^{\mu}(\tau), \label{eq:4velocity}
\end{equation} 
where $x^\mu$ is the four-position and $\tau$ is the proper time, the second term on the right side of the equation represents the stochastic contribution, being $\sigma$ the intensity of the noise and $\xi^\mu(\tau)$  a random variable with statistical properties of a Gaussian White Noise (GWN) with zero mean value $\langle \xi^{\mu}(\tau)\rangle=0$ and correlation function $\langle\xi^\mu(\tau_1)\xi^\nu(\tau_2)\rangle= \delta^{\mu\nu} \delta(\tau_2-\tau_1)$. The stochastic term causes particles to deviate from classical deterministic paths, resulting in erratic (Brownian-like) motion.

In the context of the theory of stochastic processes \cite{jacobs2010stochastic}, Eq. (\ref{eq:4velocity}) can also be written in differential notation as
\begin{equation}
    dx^\mu=\mathcal{U}^\mu d\tau+ \sqrt{2\sigma} \,d\hat{W}^\mu(\tau), \label{SDE}
\end{equation}
where $\hat{W}^{\mu}(\tau)$ is called the Wiener process such that 
$d\hat{W}^{\mu}(\tau)=\xi^{\mu}(\tau) d\tau$  which also satisfies the same statistics as the GWN $\xi^{\mu}(\tau)$ (see Appendix for the standard definition of a Wiener process in connection with a GWN), because
\begin{equation}
\langle d\hat{W}^\mu(\tau) \rangle=0, \qquad 
    \langle d\hat{W}^\mu(\tau) d\hat{W}^\nu(\tau') \rangle= \delta^{\mu\nu}\delta(\tau-\tau')  d\tau d\tau'. \label{mean-variance}
\end{equation}
The $\delta(\tau-\tau')$ is the Markovian property, meaning that the processes at different times are independent, however, the variance when $\tau=\tau'$ is $\langle (d\hat W^\mu(\tau))d\hat{W}^\nu(\tau)\rangle= \delta^{\mu\nu} d\tau$. In this case, it is said that the Wiener process is a Markovian and Gaussian stochastic process, the same as the dynamic variable $x^\mu(\tau)$ given by Eq. (\ref{SDE}). Let's recall that in the particle's reference frame, $d\tau=N(t, x^i )dt$.

In Nelson's model, time-reversal is treated through two distinct Markov processes: one representing forward time evolution $\hat{W}_{+}^{\mu}(\tau)$ and the other representing backward time evolution $\hat{W}_{-}^{\mu}(\tau)$. The mathematical formulation allows predictions about both future and past states based on the current state. So, introducing the generalized forward and backward stochastic differential equations for eq. (\ref{SDE})
\begin{align}
dx_{+}^{\mu}=\mathcal{U}_{f}^{\mu} d  \tau + \sqrt{2 \sigma} d \hat{W}_{+}^{\mu}(\tau),\label{eq:Forward SDE}\\
dx_{-}^{\mu}=\mathcal{U}_{b}^{\mu} d  \tau + \sqrt{2 \sigma} d \hat{W}_{-}^{\mu}(\tau).\label{eq:Backward SDE}
\end{align}
Where $d\hat{W}^\mu_+(\tau)$ and $d\hat{W}^\mu_-(\tau)$ satisfy the properties of Gaussian white noise described before, the zero mean value
\begin{equation} 
    \langle d\hat{W}^\mu_+(\tau) \rangle=\langle d\hat{W}^\mu_-(\tau) \rangle=0, \label{mean zero}
\end{equation} 
and correlation function 
\begin{equation}
    \langle d \hat{W}^{\mu}_{\pm}(\tau) d \hat{W}^{\nu}_{\pm}(\tau')\rangle= \pm d\tau d\tau'  \delta(\tau- \tau') \delta^{\mu \nu} \label{correlation func}
\end{equation}

The drifts define a generalized current velocity given by the forward average
\begin{equation}
    \pi^\mu=\frac{\mathcal{U}^\mu_f+\mathcal{U}^\mu_b}{2}, \label{average 1}
\end{equation}and the stochastic velocity corresponding to the backward average of the drifts
\begin{equation}
    u^\mu=\frac{\mathcal{U}^\mu_f-\mathcal{U}^\mu_b}{2}.\label{average 2}
\end{equation}

To obtain the explicit form of the two velocities, we start with the wavefunction transformation\footnote{In this section and the following, we will consider the appropriate units}
\begin{equation}
\Phi(t, \mathbf{x}) = \sqrt{n} e^{\frac{i}{\hbar}(S - \frac{\omega_0 \hbar}{c} x^0)}, \label{eq:madelungUnits}
\end{equation}here we consider $x^0=ct,$ this equation is the same as (\ref{eq:madelung}) but natural units are no longer being considered. It is important to note that now $\Phi$ stands only as a wave function and not a boson particle, or a scalar field.

The wavefunction can also be written in its WKB form as
\begin{equation}
   \Phi=e^{i\mathcal{S}/\hbar}, \label{WKB}
\end{equation}with $\mathcal{S}$ the complex action. Since $\mathcal{S}=-i\hbar\ln \Phi,$ then the complex impulse $\mathcal{P}=mw^\mu$, and the complex velocity field  $w^\mu=\frac{\nabla^\mu \mathcal{S}}{m}$ are defined in terms of (\ref{WKB}) as following
\begin{equation}
    \mathcal{P}=-i\hbar\nabla^\mu \ln{\Phi},
\end{equation}
and
\begin{equation}
    w^\mu=-\frac{i\hbar}{m}\nabla^\mu(\ln \Phi).
\end{equation}When considering the Madelung representation (\ref{eq:madelungUnits}), the complex action is related to the phase and probability density by
\begin{equation}
    \mathcal{S}=S - \omega_0 \hbar t - \frac{i\hbar}{2}\ln n,
\end{equation}and the velocity $w^\mu=\frac{\nabla^\mu \mathcal{S}}{m}$ now is defined as
\begin{equation}
    w^\mu= \frac{1}{m}\left(\nabla^\mu S - \omega_0 \hbar \nabla^\mu t -  \frac{i\hbar}{2}\nabla^\mu(\ln n) \right). \label{total velocity}
\end{equation} Here we can identify two velocities, one depending only on the phase
\begin{equation} \label{Pi sin e}
    \pi^\mu= \frac{\nabla^\mu S}{m} - \frac{\omega_0 \hbar}{m} \nabla^\mu t,
\end{equation}the second velocity, depending on the amplitude, is identified with a stochastic velocity  defined as
\begin{equation}\label{u relativista}
    u^\mu=\frac{\hbar}{2m}\nabla^\mu(\ln n).
\end{equation}
When the electromagnetic field is considered minimally coupled, we can add the gauge vector field associated with the Maxwell 4-potential to $\pi^\mu,$ such that we recover
\begin{equation}
    \pi^\mu= \frac{\nabla^\mu S}{m} + \frac{e A^\mu}{m} - \frac{\omega_0 \hbar}{m} \nabla^\mu t = v^\mu - \frac{\omega_0 \hbar}{m} \nabla^\mu t, \label{pi relativista}
\end{equation}
where $v^\mu=\frac{1}{m}\left( \nabla^\mu S + eA^\mu \right)$ is recalled. Note that by direct calculation we have that the flux (\ref{eq:flux}) $J_\mu=\frac{n}{m}\pi_\mu$, this also justifies the definition of the velocity $\pi_\mu$, since the fluxes are the density times the velocity.
Finally the total velocity (\ref{total velocity}) can be rewritten as
\begin{equation}
    w^\mu=\pi^\mu -i u^\mu, \label{total velocity w}
\end{equation}this velocity is also reported in \cite{chavanis2017derivation, zastawniak1990relativistic}. 
Here we can see the form of the diffusion constant
\begin{equation}
    \sigma=\frac{\hbar}{2m},
\end{equation}since $u^\mu \equiv \sigma\nabla^\mu(\ln{n}).$ The differential stochastic equations (forward and backward) in terms of velocities $\pi^\mu$ and $u^\mu$ can be obtained using the relations (\ref{average 1}) and (\ref{average 2}), where 
\begin{equation}
    \mathcal{U}_{f}^{\mu}=\pi^\mu + u^\mu, \quad \text{and}\quad\mathcal{U}_{b}^{\mu}=\pi^\mu - u^\mu,
\end{equation}
then equations (\ref{eq:Forward SDE}) and (\ref{eq:Backward SDE}) read
\begin{align}
dx_{+}^{\mu}=(\pi^\mu + u^\mu) d  \tau + \sqrt{2 \sigma} d \hat{W}_{+}^{\mu}(\tau),\label{eq:Forward SDE complete}\\
dx_{-}^{\mu}=(\pi^\mu - u^\mu) d  \tau + \sqrt{2 \sigma} d \hat{W}_{-}^{\mu}(\tau).\label{eq:Backward SDE complete}
\end{align}These representations could be helpful to obtain numerical simulations similar to the ones reported in \cite{Carosso_2024}, but in curved spacetime.
\\
On the other hand, it is easy for a given stochastic differential equation to obtain its corresponding Fokker-Planck equation (FPE) for the probability density, which can be written as a continuity equation \cite{risken1996fokker}. In the case of forward Eq. (\ref{eq:Forward SDE complete}) and backward (\ref{eq:Backward SDE complete}), the respective  FPEs are given by
\begin{equation}
    \nabla_\mu\left[(\pi^\mu + u^\mu) n\right] - \sigma \nabla_\mu \nabla^\mu n = 0,\label{F-P Forward}
\end{equation}
and 
\begin{equation}
    \nabla_\mu\left[(\pi^\mu - u^\mu) n\right] + \sigma \nabla_\mu \nabla^\mu n = 0, \label{F-P Backward}
\end{equation}
recalling that both processes lead to the same probability density $\Phi\Phi^\dagger=n$.

The equivalence can be seen when both equations are subtracted, then we get 
\begin{equation}
    2\nabla_\mu[(u^\mu)n]-2\sigma\nabla_\mu \nabla^\mu n = 0,
\end{equation}
and integrating this last equation the stochastic velocity (\ref{u relativista}) is recovered
\begin{equation}
     u^\mu=\sigma\frac{\nabla^\mu n}{n}.
\end{equation}
Now, by adding these equations we obtain an expression just for the systematic velocity
\begin{equation}
    2 \nabla_\mu(\pi^\mu n)=0, \label{continuity}
\end{equation}substituting (\ref{pi relativista}) we obtain
\begin{equation}
    \nabla_\mu(n v^\mu)-\frac{\omega_0}{m} \left( \nabla^0 n + n \square t \right)=0, 
\end{equation}which is exactly the continuity equation (\ref{Continuity Tona}), but obtained from a stochastic formalism.\\
Given the general velocity, it is possible to find the equation of motion for stochastic particles by fixing the total acceleration as
 \begin{equation}
     ma^\mu \equiv m\frac{dw^\mu}{d\tau}=f^\mu. 
 \end{equation}
 where electromagnetic contributions through speed $\pi^\mu$ are taken into account, and $f^\mu$ is the generalized force. However, since a stochastic differential equation is fulfilled, it is necessary to consider Ito's rule \cite{jacobs2010stochastic}. To obtain a differential operator for this kind of trajectories, we start considering the Taylor series expansion for a function $g(x)$ truncated at second-order
\begin{equation}
 dg(x) = \nabla_\mu g(x)dx^\mu + \frac{1}{2} \nabla_\mu \nabla_\nu g(x) dx^\mu dx^\nu
 \end{equation}
 where $\nabla_\mu$ stands for the total derivative, that is the covariant derivative. For this expansion, using the stochastic differential equation (\ref{eq:Forward SDE}) we have \\
 \begin{align}
     d_+g(x^\mu)&= \nabla_\mu g(x^\mu) \left[(\pi^\mu + u^\mu) d  \tau + \sqrt{2 \sigma} d \hat{W}_{+}^{\mu}(\tau) \right] \\ 
     &+ \frac{1}{2} \nabla_\mu\nabla_\nu g(x^\mu)\left[(\pi^\mu + u^\mu) d  \tau + \sqrt{2 \sigma} d \hat{W}_{+}^{\mu}(\tau) \right]\left[(\pi^\nu + u^\nu) d  \tau + \sqrt{2 \sigma} d \hat{W}_{+}^{\nu}(\tau) \right], \notag
 \end{align}
Here we use the important results of stochastic processes, normally all the infinitesimal increments of second order should be considered zero, but we keep the factor $ d\hat{W}_{+}^{\mu}(\tau)d \hat{W}_{+}^{\nu}(\tau),$ and the rest null ($d \tau d \hat{W}^{\mu}_{+}$ and $(d \tau)^2$), since we will use the Ito Rule for independent Wiener noises given by $\langle (d\hat W^\mu(\tau))d\hat{W}^\nu(\tau)\rangle= \delta^{\mu\nu} d\tau$, then we have
\begin{align}
     d_+g(x^\mu)&= \left[(\pi^\mu + u^\mu) \nabla_\mu g(x^\mu) d  \tau \right]+ \sqrt{2\sigma} \nabla_\mu g(x^\mu) d\hat{W}_{+}^{\mu}(\tau) \notag\\ 
     &+  \sigma \nabla_\mu\nabla_\nu g(x^\mu) d\hat{W}_{+}^{\mu}(\tau)d \hat{W}_{+}^{\nu}(\tau)
\end{align} 
The differential forward operator is defined as the expectation value taken over all possible events or realizations \cite{Nelson_1966} of the background randomness characterized by the noise term $\xi^\mu(\tau)$. Hence, the average over the ensemble of particles that reproduce all potential trajectories, and using the statistical properties of the aforementioned Wiener process 
 and Ito's rule we have
 \begin{eqnarray}
     \hat{D}_+g(x^\mu)&=&\left< \frac{d_{+}}{d\tau} g(x^\mu) \right>\\
     &=&\left[(\pi^\mu + u^\mu) \nabla_\mu +  \sigma \nabla_\mu\nabla^\mu \right]g(x^\mu),
 \end{eqnarray}
 When considering the backward stochastic process (\ref{eq:Backward SDE}), the backward total derivative is defined similarly using the properties for the backward Wiener process
\begin{eqnarray}
     \hat{D}_{-}g(x^\mu)&=&\left< \frac{d_{-}}{d\tau} g(x^\mu) \right>\\
     &=&\left[(\pi^\mu - u^\mu) \nabla_\mu - \sigma \nabla_\mu\nabla^\mu \right]g(x^\mu),
 \end{eqnarray}
 These two differential operators can be combined to define the systematic derivative
 \begin{equation}\label{Dc rel}
     \hat{D}_c=\frac{\hat{D}_+ + \hat{D}_-}{2}=\pi^\mu\nabla_\mu,
 \end{equation} and the stochastic derivative
 \begin{equation}
     \hat{D}_s=\frac{\hat{D}_+ - \hat{D}_-}{2}=u^\mu\nabla_\mu + \sigma \nabla^\mu\nabla_\mu.
 \end{equation}
 Since we are considering quantum particles in motion, the vicinity spacetime regions to the particles are locally flat so $\nabla_\nu x^\mu=\delta^\mu_\nu$, and it is straightforward to see that
 \begin{equation}
     \hat{D}_c x^\mu = \pi^\mu,
 \end{equation} and 
 \begin{equation}
     \hat{D}_s x^\mu = u^\mu.
 \end{equation} To recover the most general differential operator, we consider a linear combination for the total derivative defined as
\begin{equation}
    \hat{\mathcal{D}}=\hat{D}_c +\kappa \hat{D}_s,
\label{eq:total derivative}
\end{equation}where $\kappa$ is an arbitrary value. If we apply $\hat{\mathcal{D}}$ to $x^{\mu}$, we recover the total velocity $\omega^{\mu}$ as
\begin{equation}
\hat{\mathcal{D}}x^{\mu}= (\hat{D}_c -i \hat{D}_s) x^{\mu} = \pi^{\mu} -i u^{\mu}=\omega^{\mu}.
\label{eq4:5}
\end{equation}
Comparing equation (\ref{eq4:5}) with (\ref{total velocity w}), it is observed that for both velocities to be equal $(w^{\mu}=\omega^{\mu})$, consequently $\kappa=-i$.

The generalization for the KG equations is obtained when considering the total acceleration given by
\begin{equation}
    a^\mu = \hat{\mathcal{D}}w^{\mu} =(\hat{D}_{c}-i \hat{D}_{s})(\pi^{\mu}-i u^{\mu})=\hat{D}_{c}\pi^{\mu}-\hat{D}_{s}u^{\mu}-i(\hat{D}_{c}u^{\mu}+\hat{D}_{s}\pi^{\mu}) , \label{total acceleration}
\end{equation}where the different accelerations can be defined as the combination of the differential operators with the different velocities
\begin{align}
a^\mu_{cc} & = \hat{D}_c \hat{D}_c x^\mu = \hat{D}_c \pi^\mu, \quad a^\mu_{ss}  = \hat{D}_s \hat{D}_s x^\mu = \hat{D}_s u^\mu,   \\
a^\mu_{cs} & = \hat{D}_c \hat{D}_s x^\mu = \hat{D}_c u^\mu, \quad
a^\mu_{sc}  = \hat{D}_s \hat{D}_c x^\mu = \hat{D}_s \pi^\mu. \label{eq:Acceleration rel}
\end{align}
To obtain explicit expressions for the accelerations we use the following identity
\begin{equation}
    \nabla_\alpha\pi^\mu=\nabla^\mu\pi_\alpha+\frac{e}{m}F_\alpha^\mu, \label{Lorentz identity}
\end{equation}
with the Maxwell tensor defined as
\begin{equation}
    F^\mu_\alpha=\nabla^\mu A_\alpha-\nabla_\alpha A^\mu.
\end{equation}
In that case, it is straightforward to see that
\begin{eqnarray}
    \hat{D}_c\pi^\mu&=&\frac{1}{2}\nabla^\mu(\pi_\alpha\pi^\alpha) + \frac{e}{m} F^{\mu\alpha} \pi_\alpha, \label{eq:Dcc}\\
\hat{D}_s\pi^\mu&=&\sigma\left(\nabla^\mu\nabla_\alpha\pi^\alpha+\frac{\nabla_\alpha n}{n}\nabla^\mu\pi^\alpha\right)\nonumber\\
&+& \frac{e}{m} \left(  F^{\mu\alpha} u_\alpha + \sigma \nabla_\alpha F^{\mu\alpha} \right) \label{eq:Dsc}
\end{eqnarray}
In the same way, we obtain
\begin{eqnarray}
\hat{D}_cu^\mu&=&\sigma\pi_\alpha\nabla^\alpha\left(\frac{\nabla^\mu n}{n}\right) \label{Dcs}\\
     \hat{D}_su^\mu&=& 2\sigma^2\nabla^\mu \left( \frac{\square \sqrt{n}}{\sqrt{n}} \right).\label{eq:QuantumAcc}
\end{eqnarray}
Equation (\ref{total acceleration}) contains all the information that causes the dynamics of the particles, including those forces responsible for the stochastic trajectories. The total net force $f^{\mu}$ is decomposed as $f^{\mu}=f_{+}^{\mu}+if_{-}^{\mu}$. Taking the linearity between forces and accelerations, we can split the real and imaginary parts as
\begin{align} 
     m(\hat{D}_{c}\pi^{\mu}-\hat{D}_{s}u^{\mu})= f_{+} \label{eq:real acc},\\
    -m(\hat{D}_{c}u^{\mu}+\hat{D}_{s}\pi^{\mu})=f_{-}  \label{eq: imaginary acc}.
\end{align}
For the first equation (\ref{eq:real acc}), we have a purely deterministic acceleration $a_{cc}=\hat{D}_{c}\pi^{\mu}$ and a stochastic acceleration $a_{ss}=\hat{D}_{s}u^{\mu},$ using (\ref{eq:Dcc}) and (\ref{eq:QuantumAcc}) we get
\begin{equation}
    \frac{1}{2}\nabla^\mu(\pi_\alpha\pi^\alpha) + \frac{e}{m} F^{\mu\alpha} \pi_\alpha - 2\sigma^2\nabla^\mu \left( \frac{\square \sqrt{n}}{\sqrt{n}} \right) = \frac{1}{m}f^\mu_{+}. \label{62}
\end{equation}We can identify $f_+$ as external forces, that is $f_+= F^{\mu} + F_E^\mu,$ where $F_E^\mu=eF^{\alpha\mu}\pi_\alpha$ refers to the Lorentz force due to the external electromagnetic potential $A^\mu$, and $F^{\mu}=-m\nabla^{\mu}\mathcal{A}(n)$ being a force that measure the variations of the density of the scalar field as it is associated with an arbitrary potential of the form $V=2m^{2}\mathcal{A}n.$ Since the Lorentz force on the left-side emerges from the identity (\ref{Lorentz identity}), then both terms cancel on both sides, and we have
\begin{equation}
    \nabla^{\mu}\left[\dfrac{1}{2} \left(\pi_{\alpha}\pi^{\alpha}\right)-2\sigma^{2}\left( \dfrac{\Box \sqrt{n}}{\sqrt{n}}\right)+\mathcal{A}\right]=0, \label{acc=0}
\end{equation}
Observe that if we neglect any external force $f_{+}=0,$ hence the electromagnetic contributions through speed $\pi_\mu$ is also zero, and the stochastic contribution in (\ref{eq:real acc}) we have 
\begin{equation}\label{eq:geo}
\hat{D}_c \pi^\mu = \pi_\alpha\nabla^\alpha \pi^\mu=\nabla_{\pi} \pi^\mu,
\end{equation}
in the absence of forces $\hat{D}_c\pi^\mu=0$ and
(\ref{eq:geo}) becomes the definition of the vector $\pi^\alpha$ whose integral curve is just that of a geodesic. This agrees with the definition of the trajectories around geodesics, when the stochastic behavior is not considered, then the geodesics rule the trajectories instead of stochastic trajectories.\footnote{Note that in this case we consider the definition of $\pi^\mu$ without the electromagnetic contribution given in eq. (\ref{Pi sin e}).}

For the imaginary part (\ref{eq: imaginary acc}), using (\ref{Dcs}) and (\ref{eq:Dsc}), we have the following
\begin{equation}
    \nabla^{\mu}\left[\sigma \nabla_{\alpha}\pi^{\alpha}+ u_{\alpha} \pi^{\alpha} \right] + \frac{e}{m} \left(  F^{\mu\alpha} u_\alpha + \sigma \nabla_\alpha F^{\mu\alpha} \right)=\frac{f_{-}}{m}.
\label{eq4:8}
\end{equation} 
It is straightforward to see that using (\ref{u relativista}) and (\ref{pi relativista}) the terms in the brackets is 
\begin{equation}
    \sigma\nabla_\alpha\pi^\alpha + u_\alpha\pi^\alpha = \sigma n \nabla_\alpha \pi^\alpha + \sigma \pi^\alpha \nabla_\alpha n = \frac{\sigma}{n} \nabla_\alpha(n \pi^\alpha)  =0,
\end{equation} 
which coincides with the continuity equation (\ref{continuity}), therefore it is zero. And, the force $f_-$ can be read as $f_-=e(\sigma\nabla_\alpha F^{\alpha\mu}+u_\alpha F^{\mu\alpha})=\frac{e}{n}\nabla_\alpha (nF^{\alpha \mu}).$ Note that if the electromagnetic field is null, then $f_-=0,$ and eq. (\ref{eq: imaginary acc}) is just the continuity equation.

Integrating equation (\ref{acc=0}), we obtain
\begin{equation}
    \frac{1}{2}\pi_\alpha\pi^\alpha+\mathcal{A}+C-2\sigma^2\frac{\Box\sqrt{n}}{\sqrt{n}}=0
\end{equation}
which is the equation (\ref{eq:Navier}), with $C$ an integration constant. It is now evident that the field equation of a quantum particle immersed in fluctuating and arbitrary curved spacetime is identified as the KG equation. Notably, in this context, $\Phi$ represents a function determining the stochastic trajectory of a quantum particle due to the surrounding influence rather than a bosonic particle or a scalar field.

If we fix the force $f_+= F^{\mu} + F_E^\mu,$ writing eq. (\ref{62}) in terms of the velocity $v^\mu,$ using the property (\ref{Lorentz identity}) we have
\begin{equation}\label{eq:NavierStokes}
\begin{split}
     &v_\mu \nabla^\mu v_\alpha - \frac{\omega_0 \hbar}{m} \nabla^0 v_\alpha - \frac{\omega_0^2 \hbar^2}{2m^2} \nabla_\alpha \left( \frac{1}{N^2} \right) - \frac{\hbar^2}{2 m^2} \nabla_\alpha \left( \frac{\square \sqrt{n}}{\sqrt{n}} \right) \\ & =-\nabla_\alpha \mathcal{A} - \frac{e}{m} \left( F^\mu_\alpha v_\mu -\frac{\omega_0 \hbar}{m} F^0_\alpha \right).
\end{split}
\end{equation}Which is exactly the Euler equation presented in \cite{Matos_2019}, but with units. Once integrated this equation, the Hamilton-Jacobi (\ref{H-J Tona}), is recovered.  
Both equations for the hydrodynamic representation of the KG equation in curved spacetime are obtained.

    
The equation (\ref{eq:NavierStokes}) is a very useful expression, as long as we can rewrite it as
\begin{equation}
    v_\mu \nabla^\mu v_\alpha - \frac{\omega_0 \hbar}{m}\nabla^0 v_\alpha =F^E_\alpha + F^Q_\alpha + F^G_\alpha + F^n_\alpha,
\end{equation}where the forces correspond to, the gravitational
\begin{equation}
    F^G_\alpha=\nabla_\alpha \left( \frac{\omega_0^2 \hbar^2}{2m^2 N^2}  \right),
\end{equation} the quantum
\begin{equation}
    F^Q_\alpha=\frac{\hbar^2}{2 m^2} \nabla_\alpha \left( \frac{\square \sqrt{n}}{\sqrt{n}} \right), \label{eq:QuantumForce}
\end{equation}a generalized Lorentz force
\begin{equation}
    F^E_\alpha= - \frac{e}{m} \left(  \nabla^\mu A_\alpha-\nabla_\alpha A^\mu\right)\pi_\mu,
\end{equation}
and a force that is given by the derivative of an arbitrary potential $\mathcal{A}$ associated with the self-interaction potential $V(\Phi, \Phi^\dagger)$ 
\begin{equation}
    F^n_\alpha=-\nabla_\alpha \mathcal{A}.
\end{equation} The parametrization of the potential $V$ through $\mathcal{A}$ is useful as we want to measure the spatial and temporal variations of $n$, for example when we consider the potential for the mass of the scalar particle encoded in the quadratic term of the potential $V(\Phi)=m^2\Phi\Phi^\dagger,$ the corresponding parametrization is given by $\mathcal{A}=1,$ and the force $F^n_\alpha=0.$
In \cite{Matos_2019} the potential considered introduce scalar self-interactions by using the double-well (Mexican-hat) self-interacting potential given by $V=m^2\Phi\Phi^\dagger + \frac{\lambda}{2} \left(\Phi\Phi^\dagger\right)^2,$ in this case $\mathcal{A}=1 + \frac{\lambda}{2m^2}n,$ and $F^n_\alpha=-\nabla_\alpha\frac{\lambda n}{2m^2}.$ 

The fact that these forces arise from just the generalized accelerations is in accordance with general relativity, and unlike the classical formulation, we do not have to plug the forces by hand.

In what follows we stablish the connection between the KG equation and the Schr\"odinger equation. To do so, we perform the transformation 
\begin{equation}
    \Phi=\Psi\exp(i \frac{m}{\hbar c}x^0),
\end{equation}in the KG, where $\Psi=\Psi(x^\mu)$ is also a complex function and $x^0=ct$ is again the evolution parameter of the ADM metric, if we neglect the electromagnetic interaction the D´Alembertian operator reduces to $\Box=\nabla_\mu\nabla^\mu$. With this transformation is easy to see that the KG equation (\ref{eq:KG}) transforms into 
\begin{equation}
    ic\nabla^0\Psi-\frac{\hbar}{2m}\Box\Psi + \frac{ic^2}{2} \Box t \Psi + \frac{m}{\hbar}\mathcal{A}\Psi - \frac{m}{2\hbar}\left(\frac{1}{N^2}\right)\Psi=0.
\end{equation}
In the Newtonian limit, the function $N^2=1$ in the ADM metric, since $\nabla^0=-1/c\partial/\partial t$ then the D'Alembertian operator reduces to the Laplacian in flat space $\nabla^2$ and because the time derivative divides by $c^2$, i.e., $1/c^2\partial^2/\partial t^2\sim0$, finally $\Box t=0$. In that case, this equation reduces to 
\begin{equation}
    -i\hbar\frac{\partial\Psi}{\partial t}-\frac{\hbar^2}{2m}\nabla^2\Psi + m(\mathcal{A}-\frac{1}{2})\Psi=0
\end{equation}
In \cite{Gallegos:2019pyf} it is shown that in this limit $\mathcal{A}=V+1/2$, where V is an extra potential. With this, the previous equation reduces exactly to the Schr\"odinger equation.

This implies a new interpretation of the wave function of the Schrödinger equation, here the $\Psi$ function has the interpretation of a stochastic trajectory in a fluctuating spacetime, instead of a probabilistic interpretation. All the predictions and results of quantum mechanics remain intact, because the Schrödinger equation is still the master equation, what follows from this new interpretation is to understand quantum mechanics in a different way.

\section{Conclusion and discussion}

 In previous studies such as Ref. \cite{Matos_2019}, the relationship between the KG equation and hydrodynamic equations, namely, continuity (\ref{Continuity Tona}) and Navier-Stokes (\ref{eq:NavierStokes}), was established through the Madelung transformation. However, the derivation of the KG equation from more fundamental laws akin to those governing hydrodynamics remained a question.

Our contribution addresses this by adopting the stochastic nature of quantum fluctuations within spacetime. We demonstrate that the stochastic dynamics governing the quantum particles lead to the KG equation in arbitrary curved spacetime, and under specific circumstances, they also can be reduced to the Schrödinger equation. Remarkably, this is achieved without considering the metric stochastic, instead, the metric contributes to the stochastic term of the trajectories. 

Our findings affirm the efficacy of stochastic mechanics, particularly when considered within a Markovian process framework. This is encapsulated in a Langevin-type equation where $dW(\tau)$ satisfies the Wiener process or, equivalently, where the fluctuating force or thermal noise, denoted as $\xi(\tau)$, adheres to Gaussian white noise properties, exhibiting a zero mean value and a delta correlation function $\langle \xi(\tau)\xi(\tau^{\prime})\rangle \sim \delta(\tau-\tau^{\prime})$. Our proposal extends this success to explicitly unveil the KG equation in curved spacetime, where the associated stochastic trajectory of the particle stands as a physically tangible entity, echoing Nelson's observations for Markovian processes \cite{Nelson_2012,nelson2020dynamical}.

It's notable to mention that Nelson's book \cite{nelson2020dynamical} poses intriguing open problems, one of which involves stochastic mechanics within the formulation of general relativity. Our present contribution resolves this problem.

Equations (\ref{Continuity Tona}) and (\ref{eq:NavierStokes}), previously derived in \cite{Matos_2019}, establish a theoretical foundation for relativistic stochastic quantum mechanics. Furthermore, our work suggests that the presence of the stochastic term raises the quantum phenomena, providing a versatile method to investigate quantum mechanics in curved spacetime. Although is beyond the scope of the present investigation, this work could provide a novel treatment for quantum mechanics in presence of gravitational fluctuations, which could explain interference patterns \cite{gardiner2004quantum}, or the collapse of the wavefunction \cite{penrose1996gravity, diosi2014gravity}.
\vskip0.2cm

Acknowledgments: ESEA thanks SECIHTI-M\'exico for the doctoral grant.
This work was also partially supported by SECIHTI M\'exico under grants  A1-S-8742, 376127, 304001.

\begin{appendix}
\section{The Wiener process}

From the point of view of the theory of stochastic processes, the Wiener process was proposed to describe the Brownian motion as a real-valued continuous-time stochastic process\cite{wiener1923differential}. In applied mathematics, it is used to represent the integral of a Gaussian White Noise (GWN) and so it is useful as a model of noise in different fields of science. In physics it is used to study Brownian motion and other types of diffusion processes.  

The Wiener process defined as $W(t)$ is a continuous function of $t\ge 0$, characterized by the following properties:
\begin{eqnarray}
(i) && W(0)=0.  \nonumber \\
(ii) && {\rm All~increments}~ W(t_1)-W(t_0), \ldots, W(t_n)-W(t_{n-1}) \nonumber\\
&&{\rm for}~0=t_0<t_1<\ldots < t_{n-1}<t_n ~{\rm are~ independent}.  \nonumber \\
(iii) && {\rm Each~increment~is~normal~distributed~with~ mean~value}  \nonumber\\
&&E[W(t_{i+1})-W(t_i)]=0,~{\rm and~variance}~
Var[W(t_i+1)-W(t_i)]=t_{i+1}-t_i. \nonumber 
\end{eqnarray}
On the other hand, the statistical properties of a GWN $\xi(t)$, means that is has zero mean value $E[\xi(t)]\equiv\langle\xi(t)\rangle=0$ and correlation function $E[\xi(t_1)\xi(t_2)]\equiv\langle \xi(t_1)\xi(t_2) \rangle= \delta(t_2-t_1)$. Also, $W(t)$ can be defined as the definite integral of $\xi(t)$ as follows  \cite{Keng_2022} 
\begin{equation} \label{Wt}
W(\tau)=\int_0^\tau \xi(t) \, d\tau ,
\end{equation}
or equivalently, GWN is the time derivative of the Wiener process, such that 
\begin{equation} \label{dW} 
W'(t)\equiv \frac{d W(t)}{dt}=\xi(t), \qquad dW=\xi(t)\, dt,  
\end{equation}
Due to this definition, it is clear that $\langle W^{\prime}(t)\rangle=0$ and the correlation function $ \langle W^{\prime}(t_1) W^{\prime}(t_2) \rangle=\delta(t_2-t_1)$ or in terms of the differentials $\langle dW{\prime}(t_1) dW{\prime}(t_2) \rangle=\delta(t_2-t_1) dt_1 dt_2$; and  the variance must be $\langle (dW(t))^2\rangle=dt$. Moreover, the statistical properties of a GWN for the derivative, $W^{\prime}(t)$, are explicitly proven upon the definition of the derivative 
of a stochastic process, as shown in Ref. \cite{Keng_2022}.
It is the same as the GWN $\xi(t)$. So, a standard Langevin equation which we are interested in reads $dx/dt={\mathcal U}(t)+\sqrt{2\sigma}\,\xi(t)$, being $\xi(t)$ a GWN.  In differential notation it can be written as $dx={\mathcal U} dt+\sqrt{2\sigma} \, dW(t)$. The extension to a covariant formulation is thus defined by Eq. (\ref{SDE}) in Sec. 3. \end{appendix}

\bibliography{sn-bibliography}

\end{document}